\documentclass[preprint,nofootinbib,floats,showpacs,showkeys,aps,
floatfix]{revtex4}
\usepackage{graphicx}
\usepackage{bm}
\usepackage{longtable}
\def\bit{\begin{itemize}}
\def\eit{\end{itemize}}
\def\bnu{\begin{enumerate}}
\def\enu{\end{enumerate}}

\def\O {{{\cal O}}}

\def\nn{\nonumber }

\def\M {{{\cal M}}}

\def\lsim{\:\raisebox{-0.5ex}{$\stackrel{\textstyle<}{\sim}$}\:}

\def\ie{{\em i.e., }}

\def\nn{\nonumber }
\def\be{\begin{equation}}
\def\ee{\end{equation}}
\def\br{\begin{eqnarray}}
\def\er{\end{eqnarray}}
\def\brn{\begin{eqnarray*}}
\def\ern{\end{eqnarray*}}
\def\etc{ {\it etc}}

\def\pb {{\bf p}}

\def\lra{\leftrightarrow}

\def\bra#1{\langle #1|}
\def\ket#1{|#1 \rangle}
\def\rf#1{{(\ref{#1})}}

\def\go{\rightarrow  }
\def\etal {\emph{et al.}}

\def\I {{{\cal I}}}

\def\fot{\frac{1}{2}}

\def\etc{ {\it etc}}
\begin{document}
\title{Nonmesonic weak  decay spectra of $^{4}_\Lambda$He}
\author{E. Bauer$^{1,5}$}
\author{A. P. Gale\~ao$^2$}
\author{M. S. Hussein$^{3, 6}$}
\author{F. Krmpoti\'c$^{3,5,7}$}
\author{J. D. Parker$^{4}$}
\affiliation{$^1$Facultad de Ciencias Exactas,
 Departamento de F\'isica, Universidad Nacional de La Plata, 1900 La Plata,
Argentina,}

\affiliation{$^2$
Instituto de F\'{\i}sica Te\'orica,
Universidade Estadual Paulista, \\
Rua Pamplona 145,
01405-900 S\~ao Paulo, SP, Brazil,}

\affiliation{$^3$Departamento de F\'isica Matem\'atica, Instituto
de F\'isica da Universidade de S\~ao Paulo,
Caixa Postal 66318, 05315-970 S\~ao Paulo, SP, Brazil,}

\affiliation{$^4$ Department of Physics, Kyoto University, Kyoto 606-8502, 
Japan,}

\affiliation{$^5$Instituto de F\'isica La Plata, CONICET, 1900 La
Plata, Argentina,}

\affiliation{$^6$ Max-Planck-Institut f\"ur Physik komplexer Systeme,
N\"othnitzer Stra{\ss}e, 38, D-01187 Dresden, Germany,}

\affiliation{$^7$Facultad de Ciencias Astron\'omicas y
Geof\'isicas, Universidad Nacional de La Plata, 1900 La Plata,
Argentina.}

\date{\today}

\begin{abstract}
To comprehend the recent Brookhaven National Laboratory experiment
E788 on  $^4_\Lambda$He, we have outlined  a simple theoretical framework, 
based on the independent-particle shell model,
for the one-nucleon-induced nonmesonic weak decay spectra.
Basically, the shapes of all the spectra are tailored by the kinematics of
the corresponding phase space, depending very weakly on the dynamics, 
which is gauged here by the one-meson-exchange-potential.
In spite of the straightforwardness of the approach a 
good agreement with data is acheived. 
This might be an indication that the final-state-interactions
and the two-nucleon induced processes are not very important  in
the decay of this hypernucleus.
We have also found that the $\pi+K$ exchange potential
with soft vertex-form-factor cutoffs $(\Lambda_\pi\approx 0.7$ GeV,
$\Lambda_K\approx 0.9$ GeV), is able to account simultaneously for the 
available experimental data related to $\Gamma_p$ and $\Gamma_n$
for  $^4_\Lambda$H, $^4_\Lambda$He, and $^5_\Lambda$He.
\end{abstract}


\maketitle
The  nonmesonic weak decay   (NMWD)  of
$\Lambda$ hypernuclei, $\Lambda N\go nN$ ($N=p,n$),
is very interesting in several aspects.
First, it implies the most radical mutation
of an elementary particle when embedded in a nuclear environment:
without producing  any additional on-shell particle, as does  the
mesonic weak decay  $\Lambda \go \pi N$, the mass is changed by $176$ MeV, 
and the strangeness by $\Delta S=1$.  Second, it is the main decay
channel for medium and heavy hypernuclei.
Third,  as such  it offers the best opportunity to examine the 
strangeness-changing  nonleptonic weak interaction between hadrons.  
Fourth, it plays a dominant role in the stability of rotating neutron stars 
with respect to gravitational wave emission \cite{Jo01,Ju08}. 
Finally, with the incorporation of  strangeness, the $(N,Z)$ radioactivity 
domain is  extended to three dimensions $(N,Z,S)$. 
Therefore, the understanding of the NMWD cannot but help to advance
our knowledge of physics.

Several important experimental advances  in NMWD
 have been made in recent years,
which have allowed to establish more precise values of the
neutron- and proton-induced  transition rates  $
\Gamma_n\equiv\Gamma(\Lambda n\go nn)$
 and $\Gamma_p\equiv\Gamma(\Lambda p\go np)$,
solving in this way the long-standing puzzle of
the branching ratio  $\Gamma_{n/p}\equiv\Gamma_n/\Gamma_p$. 
They are: 1) the new
high quality measurements of  single-nucleon spectra $S_{N}(E)$,
as a function of one-nucleon energy $E_N\equiv E$ done in 
Refs.~\cite{Ki03,Ok04,Ag08,Pa07}, 
and 2) the first  measurements of the two-particle-coincidence spectra as a 
function of the sum of  kinetic energies $E_n+E_N\equiv E$, $S_{nN}(E)$, and of 
the opening angle $\theta_{nN}\equiv\theta$, $S_{nN}(\cos\theta)$, done in 
Refs.~\cite{Ok05,Ou05,Ka06,Ki06,Pa07,Bh07}.

Particularly interesting is the Brookhaven National Laboratory experiment
E788 on $^{4}_\Lambda$He, performed by Parker \etal~\cite{Pa07}, 
which highlighted that the effects of the Final State Interactions (FSI) on 
the one-nucleon induced decay, as well as the contributions of the two-nucleon
induced decays, $\Lambda NN \go nNN$,  could be very small in this case,
if any. 
Therefore one might hope that the Independent Particle Shell Model (IPSM) 
\cite{Pa97,Pa02,It02,Ba02,Kr03,Ba03} could be an adequate framework to account 
for the NMWD spectra of this hypernucleus. 
The aim of the present work is to verify this expectation.

To derive the expressions for the NMWD rates  we  start from the Fermi Golden
Rule~\cite{Ba02}. 
For  a hypernucleus with spin $J_I$ decaying to residual nuclei with spins 
$J_F$, and two free nucleons $nN$ (with  total spin $S$ and total kinetic energy
$E_{nN}=E_n+E_{N}$), the transition rate reads
\begin{equation}
\Gamma_N=2\pi \sum_{SM_SJ_FM_F} \int |\bra{\pb_n\pb_N SM_SJ_FM_F}
V\ket{J_IM_I}|^2 \delta(E_{nN}+E_R-\Delta_N) \frac{d{\bf
p}_n}{(2\pi)^3}\frac{d{\bf p}_N}{(2\pi)^3}.
 \label{1}
\end{equation}
The NMWD dynamics, contained within the  weak hypernuclear
transition potential $V$, will be described  by the one-meson exchange (OME)
model, whose most commonly used version  includes the exchange
of the full pseudoscalar ($\pi, K, \eta$) and vector
($\rho,\omega,K^*$) meson octets (PSVE), with the weak coupling
constants obtained from soft meson theorems and
$SU(6)_W$~\cite{Pa97,Du96}.
The wave functions for the kets
$\ket{\pb_n\pb_N SM_SJ_FM_F}$ and $\ket{J_IM_I}$ are assumed to be
antisymmetrized and normalized, and the two emitted nucleons $n$
and $N$ are described by plane waves.  Initial and final short
range correlations are included phenomenologically at a simple
Jastrow-like level, while the finite nucleon size effects at the
interaction vertices are gauged by monopole form
factors~\cite{Pa97,Ba02}. Moreover,
 \be
 E_R =\frac{|\bm{p}_n+
\bm{p}_N|^2}{2M(A-2)}= \frac{E_{nN} + 2\cos\theta_{nN} \sqrt{E_n
E_N}}{A - 2},
 \label{2}\ee
 is the
recoil energy of the residual nucleus, and  $\Delta_{N}\equiv
\Delta+e_{N}+e_\Lambda$ is the liberated energy, with $\Delta=M-
M_\Lambda=176$ MeV,
 and  $e_{N}$ and $e_\Lambda$  being the nucleon and hyperon
 separation energies, which were taken from Refs.~\cite{Wa71} and ~\cite{Ko06}
respectively.

Following step by step the developments done in Refs.
\cite{Ba05,Ba07,Kr08}, in connection with the asymmetry parameter, Eq.
\rf{1} can be cast in the form \br \Gamma_N&=&\frac{4}{\pi}\int
d\cos\theta \int p_N^2dp_N \int p_n^2dp_n\, \delta\left(E_{nN}+E_R -
\Delta_{N}\right)\I_N(p,P),
 \label{3}\er
where the quantity~\cite{Ba02,Kr03,Ba07,Kr08}
 \br
\I_N(p,P)&=&\sum_{ J=0} ^{J=1}F_J(N) \sum_{SlT} \M^2(pP, l SJT;N),
\label{4}\er
  depends on the spectroscopic
 factors $F_J(N)$, and on the transition matrix elements
 $\M(pP, lSJT;N)$.
Those, in turn, depend on the  c.m. and relative  momenta, which
are given in terms of the integration variables in \rf{3} by \br
{P}&=&\sqrt{(A-2)(2M\Delta_N - p_n^2 -p_N^2)},
 \label{5}\er
and
 \br
p&=&\sqrt{M\Delta_{N}- \frac{A}{4(A-2)} P^2},
 \label{6}\er
where the energy conservation condition has been used.
The correctness of Eq. \rf{3} for $N=p$ can be easily verified by
confronting  it with  the expression \cite[Eq. (3.1)]{Ba07} for
$\omega_0\equiv\Gamma_p$, and noticing that the quantity
$\sum_{\sf l L}\O(P;{\sf L}){\cal I}_0(p;j_p,{\sf l})$ in that
reference is equal to $\I_p(p,P)$ here. The nuclear matrix elements (NME), 
that govern the NMWD dynamics proper, 
are contained within the $\M$'s and depend on $P$
only indirectly via $p$ (see \cite[Eq. (B1)]{Ba07}). Moreover,
this dependence is very weak and  allows to  compute the NME's at
the fixed value of $p=\sqrt{M \Delta_{N}}$~\cite{Ba07,Kr08}. As a
consequence the NME's can be factored out of the
integrals in Eq. \rf{3} and this explains why only the transition
rates, but not the normalized spectra, significantly depend on the
intrinsic NMWD dynamics~\cite{Ba08}. Notice, however, that the
$\M$'s as a whole do strongly depend on $P$ through the
center-of-mass overlaps of the two-body wave functions.

Next, the $\delta$-function  in \rf{3} can be put in the form
 \br
\frac{A-2}{A-1}\frac{2M}{|p_n^+-p_n^-|}
\left[\delta(p_n-p_n^+)+\delta(p_n-p_n^-)\right],
\label{7}\er
 where
 \br
 p_n^\pm
&=&(A-1)^{-1}\left[-p_N\cos{\theta_{nN}}\pm
\sqrt{2\emph{}M(A-2)(A-1)\Delta_N-p_N^2\left[(A-1)^2-\cos^2{\theta_{nN}}\right]}
\right].
\nn\\
\label{8}\er 
Doing this, Eq. \rf{3} becomes
 \br
\Gamma_N&=&\frac{8M}{\pi}\frac{A-2}{A-1} \int _{-1}^{+1}d
\cos\theta_{nN}\int p_N^2 dp_N \frac{(p_n^+)^2}{|p_n^+-p_n^-|}
\left[\I_N(p,P)\right]_{p_n \to p_n^+} + ( p_n^+\lra p_n^-),
\label{9}\er where the notation $\left[\I_N(p,P)\right]_{p_n \to
p_n^+}$ indicates that $\I_N(p,P)$ is to be computed with $P$ and
$p$ given by Eqs.~\rf{5} and \rf{6} with $p_n$ replaced by
$p_n^+$. We have shown numerically that the last term in \rf{9} is
negligibly small in comparison with the first one and therefore it
will be omitted  from now on. With the simple change of variable
$p\go \sqrt{2ME}$ one finally gets
 \br
\Gamma_N&=&(A-2)\frac{8M^3}{\pi} \int _{-1}^{+1}
d\cos\theta_{nN}\int_0^{E_{N}^{max}} dE_{N}
\sqrt{\frac{E_{N}}{E_N'}}\, E_n^+\, \I_N(p^+,P^+),
 \label{10}\er
where 
 \br
E'_N&=& 
(A-2)(A-1)\Delta_{N}-E_{N} \left[(A-1)^2-\cos^2\theta_{nN}\right],
\label{11}\er
  \br
  E_n^+&=&\left[
\sqrt{E'_N} -\sqrt{E_{N}}\cos{\theta_{nN}}\right]^2(A-1)^{-2},
\label{12}\er and $P^+$ and $p^+$ are to be computed from
Eqs.~\rf{5} and \rf{6} with $p_n$ replaced by $p_n^+$. It might be
worth noticing that, while $E'_N$ does not have a direct physical
meaning,
 $E_n^+$ is the energy of the neutron that is
the decay partner of the nucleon $N$ with energy $E_N$.  The
maximum energy of integration in \rf{10} is \be
E^{max}_{N}=\frac{A-1}{A}\Delta_{N}. \label{13}\ee
This ensures that $p_n^+$, given by Eq.~\rf{8}, is real. In order
to ensure that it also be positive, as it must, one has to
enforce the condition
\begin{equation}
\sqrt{E'_N} > \sqrt{E_{N}}\cos{\theta_{nN}} \label{14}
\end{equation}
throughout the integration.

The decay rate in Eq.~\rf{3} can be rewritten in terms of energy
variables as \br \Gamma_N&=&\frac{8M^3}{\pi}\int d\cos\theta_{nN}
\int dE_N \int dE_n \sqrt{E_NE_n} \delta\left(E_{nN}+E_R -
\Delta_{N}\right)\I_N(p,P),
 \label{15}\er
and the energy-conserving $\delta$-function as
 \br
\frac{A-2}{2\sqrt{E_nE_N}}\delta\left[\cos\theta_{nN}-
C_{nN}(E_n,E_N) \right], \label{16}\er where
\begin{equation}
C_{nN}(E_n,E_N) = 
\frac{(A - 2)\Delta_{N}-(A-1)(E_n+E_N)}{2\sqrt{E_n E_N}}. 
\label{17}
\end{equation}
Thus, upon eliminating the delta, one gets 
\br
\Gamma_N &=& \frac{4M^3(A-2)}{\pi} 
\int_0^{E^{max}_{N}}dE_N\int_0^{E^{max}_{N}}dE_n \,
\I_N(p,P), \label{18}
\er 
with the constraint
\begin{equation}
-1 < C_{nN}(E_n,E_N) < +1 \label{19}
\end{equation}
to be imposed throughout the integration. Here, the variables $P$
and $p$ in $\I_N(p,P)$ can be computed from
\br
{P}&=&\sqrt{2M(A-2)(\Delta_{N}-E_n-E_N)} \label{20} \er
and Eq.~\rf{6}.

We note that in Ref. \cite{Ba08} the kinetic energy sum spectra
have been evaluated from 
\br \Gamma_N
&=&\frac{4M^3}{\pi}\sqrt{A(A-2)^3} 
\int_{E^{min}_{nN}}^{\Delta_{N}}dE_{nN} 
\sqrt{(\Delta_{N}-E_{nN})(E_{nN}-E^{min}_{nN})} \,
\I_N(p,P) ,
\label{21}\er 
with 
\be E^{min}_{nN}=\Delta_{N}\frac{A-2}{A},
\label{22}\ee
and $p$ and $P$  given by Eq. \rf{6} and 
\br {P}&=&\sqrt{2M(A-2)(\Delta_{N}-E_{nN})}. 
\label{23}\er 
Here, however, in order to be able to take one-nucleon detection energy 
thresholds into account, it is more convenient to start from Eq. \rf{18} 
rewritten in the form
\br
\Gamma_N&=&\frac{4M^3(A-2)}{\pi}\int_{E^{min}_{nN}}^{\Delta_{N}}
dE_{nN} \int_0^{E^{max}_{N}}dE_N\int_0^{E^{max}_{N}}dE_n \I_N(p,P)
\delta(E_N+E_n-E_{nN}).
\label{24}\er
To implement angular cuts, one has simply to alter the lower and/or upper 
limits in inequality \rf{19}.

The needed transition probability densities $S_N(E_{N})$, 
$S_{nN}(\cos\theta_{nN})$, and $S_{nN}(E_{nN})$ can now be obtained
by performing derivatives on $E_{N}$, $\cos\theta_{nN}$, and
$E_{nN}$ in the appropriate equation for $\Gamma_N$, namely,
Eq.~\rf{10} or Eq.~\rf{18}  for the first, Eq.~\rf{10} for the second, 
and Eq.~\rf{21} or Eq.~\rf{24} for the third one.

The  experimental data on NMWD rates in the $s$-shell are compared in 
Table \ref{Table1} with the most recent theoretical results. 
As can be seen, no calculation,  in which
the same model and the same parametrization have been employed for
all three nuclei, is capable of reproducing all the data, which
might imply that no one of them describes the full dynamics of
these processes. 
In particular, using the PSVE model~\cite{Kr03} it was not possible to account, 
either for the $^{4}_\Lambda$He, or for the $^{5}_\Lambda$He data, 
while the potentials constructed by Itonaga \etal~\cite{It02}, 
from the correlated $2\pi$ coupled to $\rho$ and/or  $\sigma$ mesons, are
conflicting with the recent $^{4}_\Lambda$He  data for
$\Gamma_{nm}$ and $\Gamma_{n/p}$~\cite{Pa07}.
The only calculation  done with the PSVE model that reproduces the 
$^{5}_\Lambda$He data is the one by  Chumillas
\etal~\cite{Ch08}, but, unfortunately, the results for the remaining two 
$s$-shell hypernuclei are not given.  
We have repeated now the calculation done previously in Ref.~\cite{Kr03} for 
the PSVE and PKE models, but  with values of  the size parameter, $b$, taken 
from Ref.~\cite{It02}:
$b( ^4_\Lambda$H)$=b(^4_\Lambda$He)$=1.65$ fm, and $b(^5_\Lambda$He)$=1.358$ fm.
In Table \ref{Table1}, they are labelled, respectively, as  P1 and P2,  and both
are very far from data.%
\footnote{It is more than evident that the value of $b$ is important in 
scaling  the magnitudes of the $\Gamma_N$. The differences between the PSVE
results shown here and those reported in Ref.~\cite{Kr03} 
arise from the values of $b$ used. 
In the latter case that value was taken to be 
$b=\sqrt{\frac{\hbar}{M\omega}}$, 
with $\hbar \omega =45A^{-1/3}-25A^{-2/3}$ MeV.
We don't know the origin of the discrepancy with Chumillas \etal~\cite{Ch08}. }
We do not know how these $b$-values have been adjusted, but they seem to be more 
realistic than those used in Ref.~\cite{Kr03}.
In fact, they are consistent with the estimate
$b=\frac{1}{2}\sqrt{\frac{2}{3}}(R_N+R_\Lambda)$,  
where $R_N$ and $R_\Lambda$ are, respectively, the root-mean-square
distances of the nucleons and the $\Lambda$ from the center of mass of the 
hypernucleus. 
This yields $b( ^4_\Lambda$H)$=b(^4_\Lambda$He)$=1.53$ fm, 
and $b(^5_\Lambda$He)$=1.33$ fm \cite{Ko06}.
The relative and the c.m. oscillator parameters are simply evaluated as 
$b_r=b\sqrt{2}$ and $b_R=b/\sqrt{2}$. 
We have also tried \cite[Eqs. (36) and (37)]{In98},
used by Inoue \etal,  but this has little influence on our results.

\begin{figure}[h]
\begin{tabular}{cc}
\includegraphics[width=0.5\linewidth,clip=]{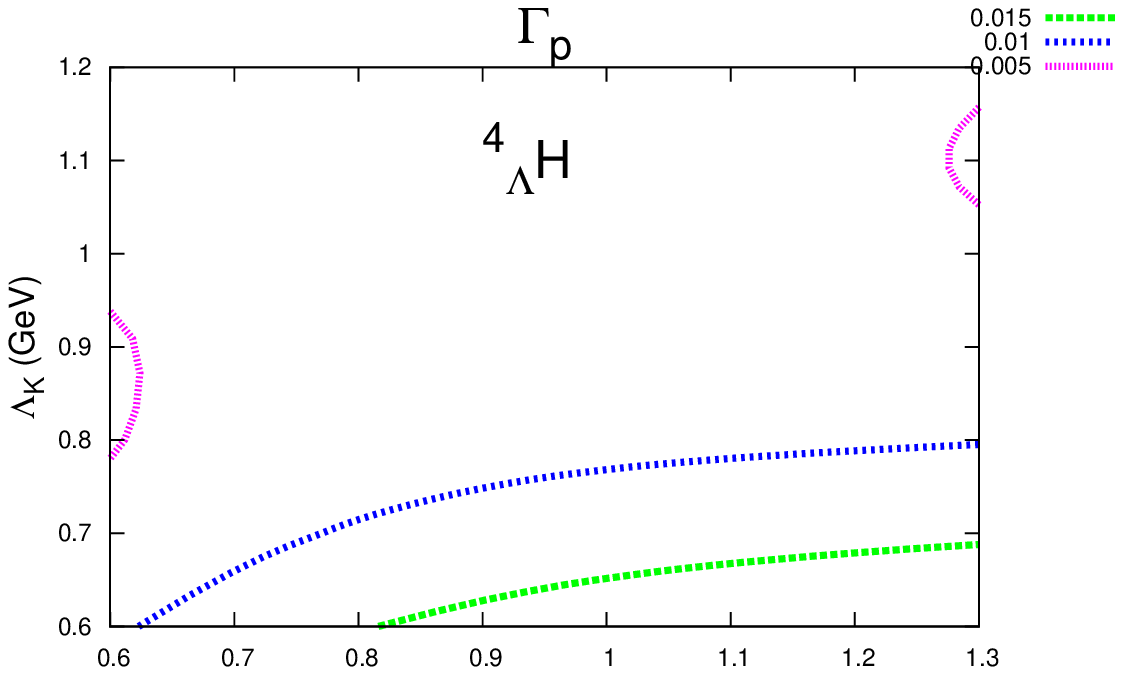} &
\includegraphics[width=0.5\linewidth,clip=]{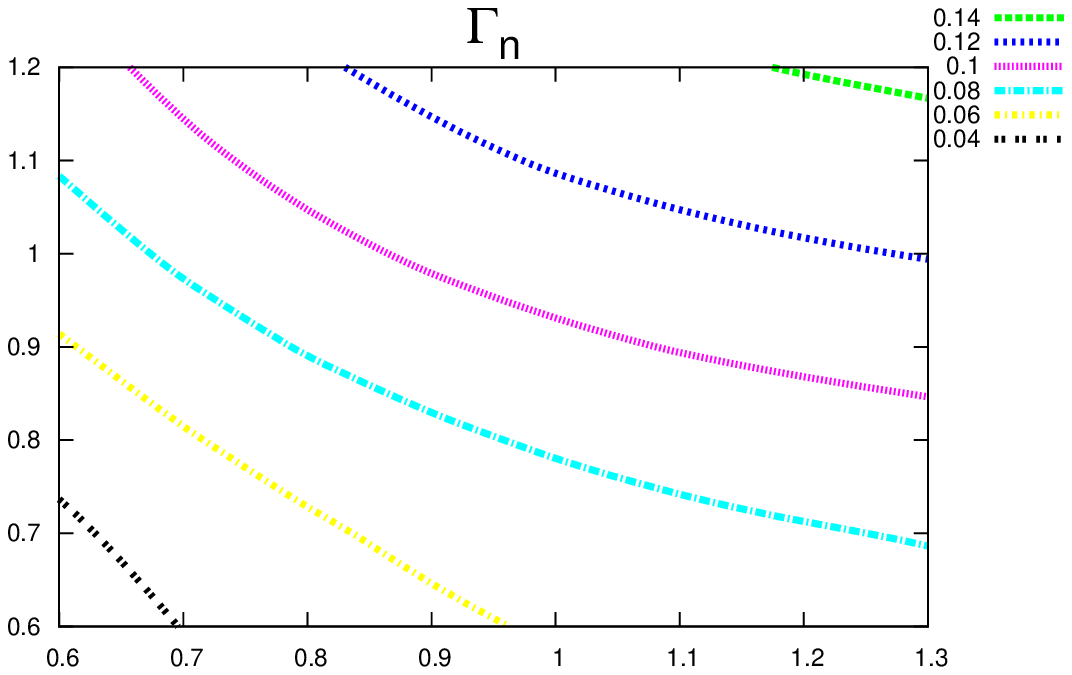} \\
\includegraphics[width=0.5\linewidth,clip=]{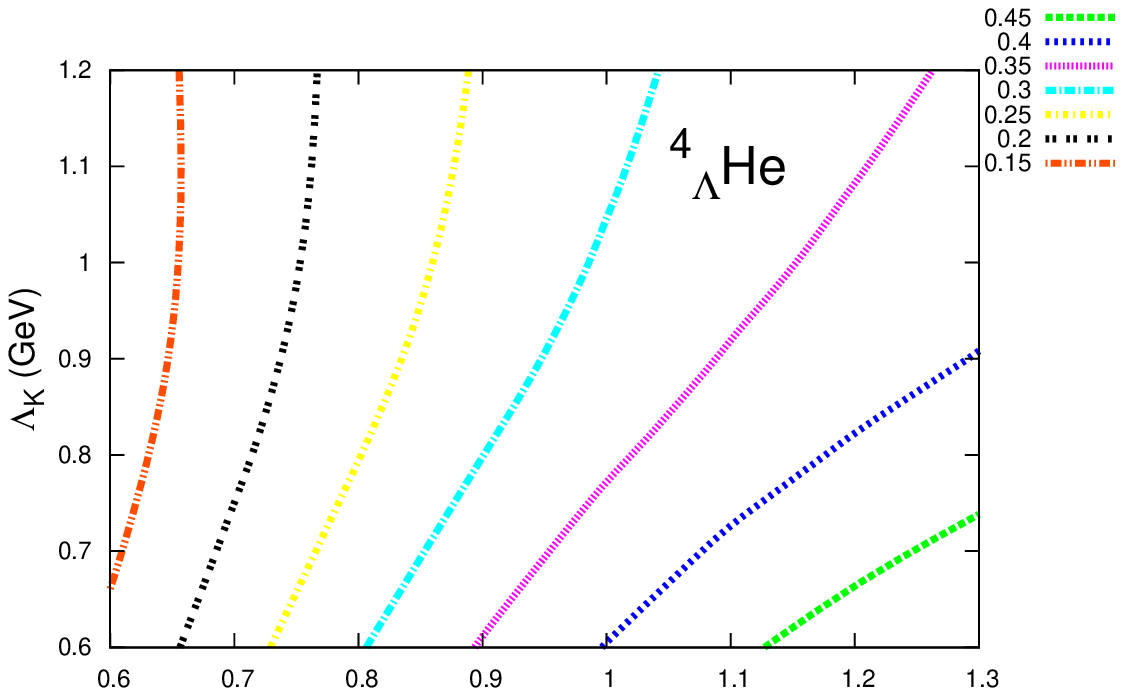} &
\includegraphics[width=0.5\linewidth,clip=]{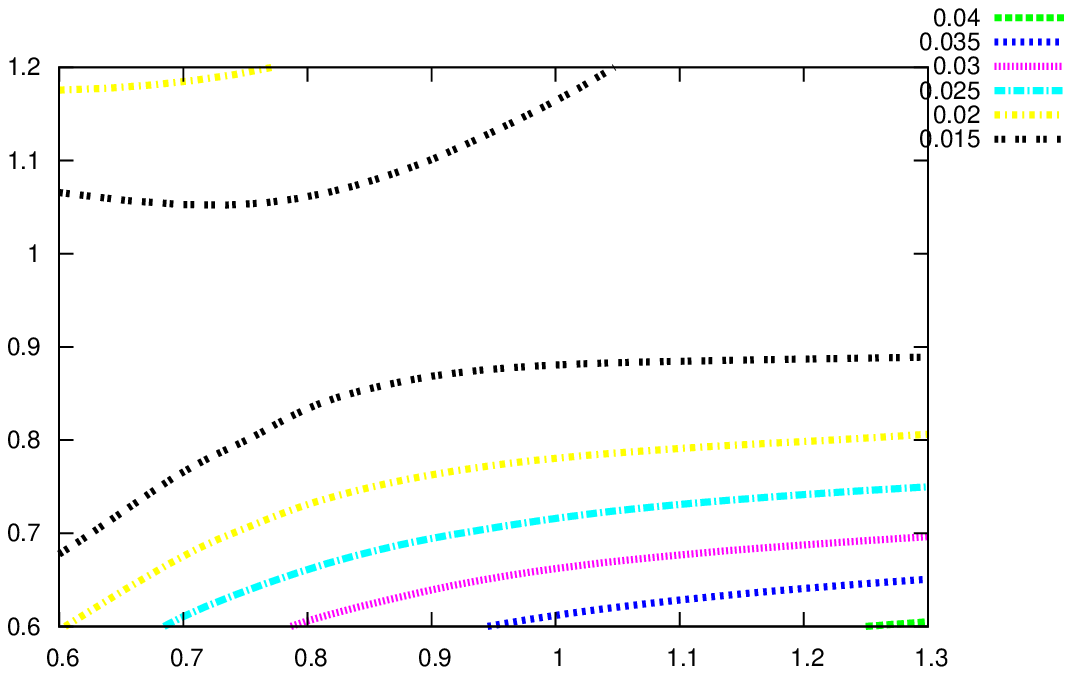} \\
\includegraphics[width=0.5\linewidth,clip=]{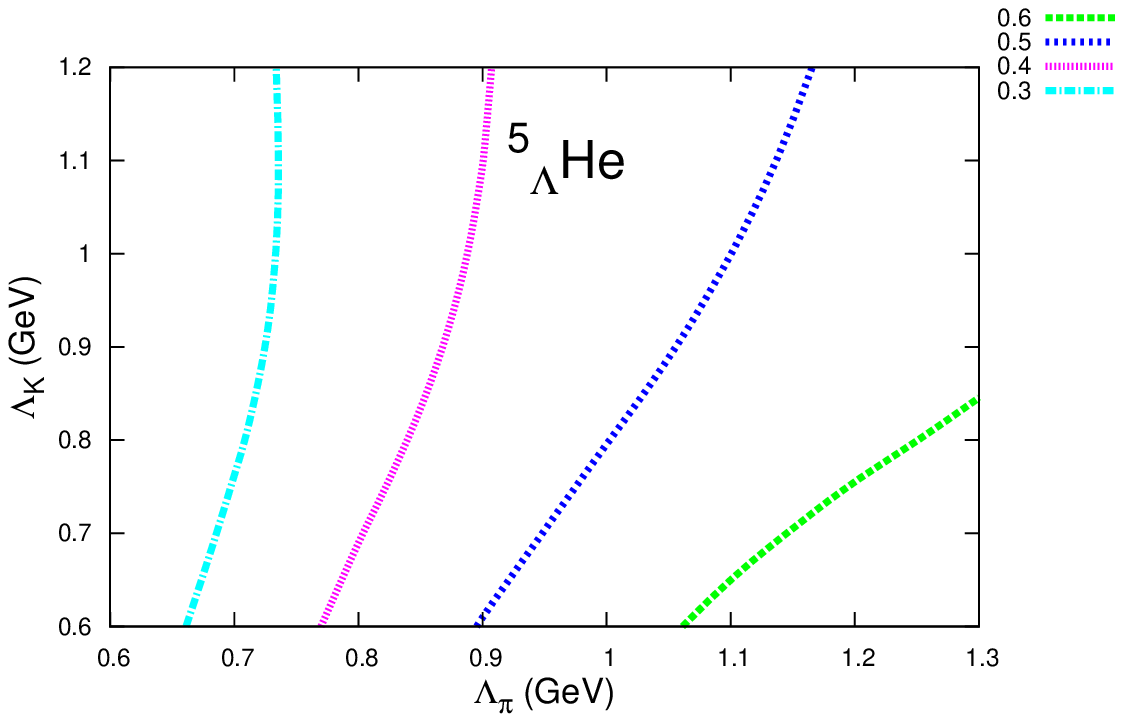} &
\includegraphics[width=0.5\linewidth,clip=]{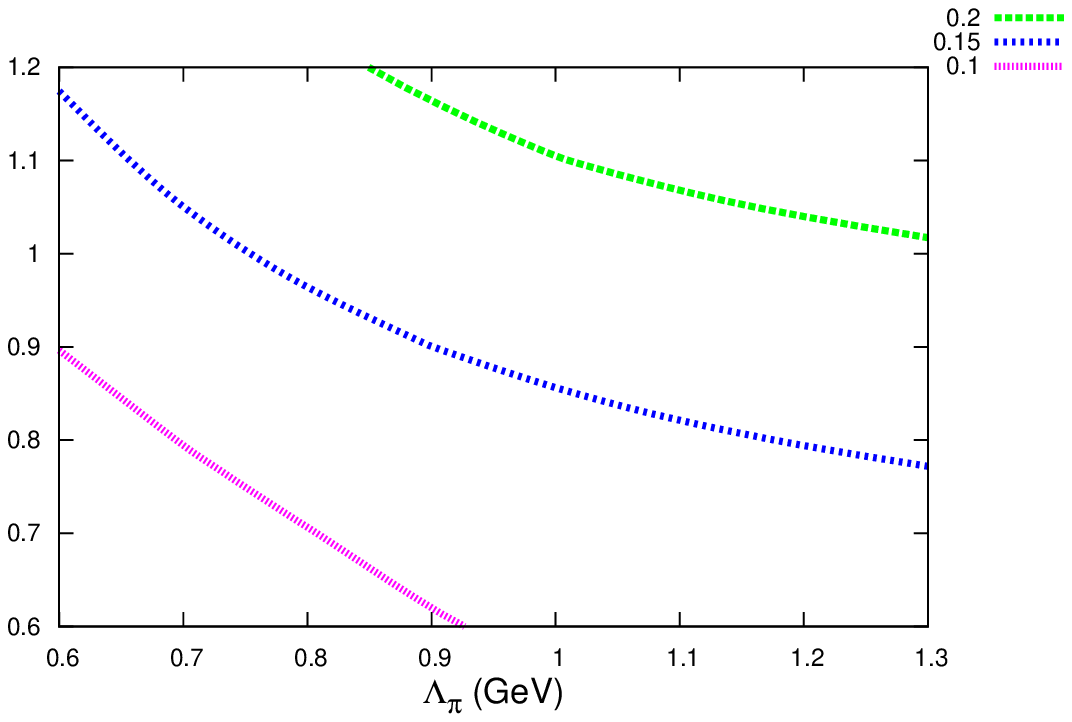}
\end{tabular}
\caption{\label{Figure1} Decay rates $\Gamma_p$ and $\Gamma_n$ of $^4_\Lambda$H,
$^4_\Lambda$He and $^5_\Lambda$He, for  fixed values of  the size
parameters, $b( ^4_\Lambda$H)$=b( ^4_\Lambda$He)$=1.65$ fm and
$b( ^5_\Lambda$He)$=1.358$ fm \cite{It02}, as a function of
pion and kaon cutoff parameters $\Lambda_\pi$ and $\Lambda_K$.}
\end{figure}

To improve the agreement we could either: 1)  add more mesons, 2)
modify the model parameters, or 3) incorporate additional degrees
of freedom.
We have  chosen the second option, trying to use the smallest number of
mesons.
The simplest possibility is, of course, the one-pion exchange potential. We have
found that for the monopole  vertex-form-factor cutoff parameter of the pion,
$\Lambda_\pi\lesssim 0.7$ GeV, and the size parameter
$b\gtrsim 1.6$ fm it is possible to account for the
$^{4}_\Lambda$He data but not for that of $^{5}_\Lambda$He.  Next,
we have examined the one-$(\pi+ K)$ exchange (PKE) model, for  fixed
values of  the size parameters $b$ mentioned above.
In Figure \ref{Figure1} is shown the  dependence of $\Gamma_{N}$
on the $\pi$ and $K$  cutoff parameters $\Lambda_\pi$ and
$\Lambda_K$.
Roughly speaking, $\Gamma_p(^4_\Lambda$He) and $\Gamma_p(^5_\Lambda$He)
depend mainly on $\Lambda_\pi$, while $\Gamma_p(^4_\Lambda$H) and
$\Gamma_n(^4_\Lambda$He) depend mainly on $\Lambda_K$ and the other two
rates depend with about equal weight on both.
The similarities and the differences in the behaviors of $\Gamma_p$
and $\Gamma_n$ for the three hypernuclei are  mainly due to the
spectroscopic factors, exhibited in ~\cite[Table 1]{Kr03}.  The
$b$-values also play a significant role. The most relevant issue
here is, however, that there is a region of rather soft
$\Lambda_\pi$ and $\Lambda_K$
where all the $\Gamma_{N}$  are reproduced fairly well.%
\footnote{To reproduce the combined effect of short-range
correlation and form factor reductions
Bennhold and  Ramos~\cite{Be92}
have used a monopole form factor with a very soft cutoff of
$\Lambda_\pi\approx 0.6$ GeV.}
In Table \ref{Table1} are shown  the results for
$\Lambda_\pi= 0.7$ GeV and  $\Lambda_K=0.9$ GeV, labelled as P3,
which we call the soft $\pi+ K$ exchange (SPKE) potential,
and which will be used in the evaluation of the NMWD spectra of 
$^{4}_\Lambda$He in what follows.
We note that they are similar to the results T3, obtained by 
Sasaki \etal~\cite{Sa02} within
the PKE model with $\Lambda_\pi= 0.8$ GeV and  $\Lambda_K=1.2$ GeV.
It is interesting to remark that the $\Delta T=\fot$ prediction
$ 
\frac{\Gamma_n(^4_\Lambda\rm{He})}{\Gamma_p(^4_\Lambda\rm{H})}=2
$ 
is quite well fulfilled for the SPKE model. Yet, the relationship
$ 
\frac{\Gamma_n(^4_\Lambda\rm{H})}{\Gamma_p(^4_\Lambda\rm{He})}=
\frac{\Gamma_n}{\Gamma_p}(^5_\Lambda\rm{He})
$ 
is satisfied only approximately. The  reason for that are the differences in 
the binding energies and the values of the $b$ parameter.

\begin{table}[h]
\caption{ The NMWD rates in the $s$-shell. \underline {A
Experimental}:  E1~\cite{Ou98}:  E2~\cite{Pa07}; E3~\cite{Ou05}; E4~\cite{Ka06},
 \underline { B) Theoretical}:
T1 - ($\pi+ DQ$)~\cite{In98};
 T2 - ($\pi + 4BPI$)~\cite{Ju01};
 T3 - (PKE)~\cite{Sa02};
 T3' - ($\pi+K + DQ$)~\cite{Sa02};
 T4 - ($\pi+2\pi/\sigma+2\pi/\rho+\omega$)~\cite{It02};
 T5 -(PSVE)~\cite{Kr03};
 T6 -  ($\pi+K +\sigma+ DQ$)~\cite{Sa05},
T7 - (PSVE)~\cite{Ch08};
T7' - (PSVE $+2\pi+2\pi/\sigma$)~\cite{Ch08}.
 \underline {C) Present Results}:
 P1 - (PSVE);
P2 - (PKE);
P3 -  (SPKE).
\label{Table1} }
\begin{tabular}{|c|cccc|cccc|cccc|}
\hline
&&&{$^4_\Lambda$H}&&&&{$^4_\Lambda$He}&&&&{$^5_\Lambda$He}&\\
\hline
&$\Gamma_p$&$\Gamma_n$&$\Gamma_{nm}$&$\Gamma_{n/p}$
&$\Gamma_p$&$\Gamma_n$&$\Gamma_{nm}$&$\Gamma_{n/p}$
&$\Gamma_p$&$\Gamma_n$&$\Gamma_{nm}$&$\Gamma_{n/p}$\\
\hline
A)         &&&&&&   &&&&&&\\
E1&&&$0.17^{+0.11}_{-0.11}$&&$0.16^{+ 0.02}_{- 0.02}$ & $0.01^{+0.04}_{-0.01}$ 
& $0.17^{+0.05}_{-0.05}$ & $0.06^{+0.25}_{-0.06}$&&&&\\
E2&&&&&$0.180^{+ 0.028}_{- 0.028}$&$\le 0.035$
&$0.177^{+0.029}_{-0.029}$&$\le 0.19$&&&&\\
E3 &&&&&&&&&&&$0.424^{+0.024}_{-0.024}$&\\
E4 &&&&&&&&&&&&$\hspace{0.1cm}0.45^{+0.14}_{-0.14}\hspace{0.1cm}$\\
\hline
B)         &&&&&&   &&&&&&\\
T1&$ 0.047 $&$0.126 $&$0.174 $&$2.66$&$ 0.214$&$0.038 $&$0.253$&$0.178$
&$ 0.421 $&$   0.206 $&$0.627 $&$0.489$\\
T2&$ 0.034 $&$0.002 $&$0.036 $&$18.2$&$ 0.030 $&$0.170 $&$0.200 $&$0.17$
&$ 0.192 $&$   0.174 $&$0.366 $&$1.10$\\
T3&$ 0.005 $&$0.067 $&$0.071 $&$14.2$&$ 0.145 $&$0.009 $&$0.155 $&$0.064$
&$ 0.207 $&$   0.097 $&$0.304 $&$0.466$\\
T3'& $ 0.030$&$ 0.157$&$    0.187 $&$5.32$&
                $ 0.214$&$ 0.004$&$    0.218$&$   0.019$&
                $ 0.304$&$ 0.219$&$    0.523 $&$   0.720$\\
T4& $ 0.040 $&$   0.088  $&$    0.128 $&$    2.17$
& $ 0.223 $&$   0.081  $&$    0.303 $&$    0.363$
& $ 0.305$&$   0.118  $&$    0.422 $&$    0.386$\\
T5& $ 0.014 $&$   0.154  $&$    0.168 $&$    10.4$
& $ 0.477 $&$   0.030 $&$    0.507 $&$   0.061$
& $ 0.461 $&$   0.148$&$    0.609 $&$   0.320$\\
T6& $0.035 $&$0.093    $&$    0.128 $&$    2.70$
& $0.165 $&$ 0.069   $&$    0.235 $&$    0.417$
& $0.253 $&$ 0.392   $&$    0.392$&$    0.548$\\
T7&&&&&&&&& $0.257 $&$ 0.122   $&$    0.474$&$    0.379$\\
T7'&&&&&&&&& $0.275 $&$ 0.114   $&$    0.415$&$    0.388$\\
\hline
C)         &&&&&&   &&&&&&\\
P1&$0.014 $&$    0.144  $&$   0.159 $&$   9.98  $&$ 0.463   $
&$0.029   $&$0.492 $&$0.062 $&$0.701 $&$0.229  $&$0.930$&$0.327$\\
P2&$0.005 $&$0.143 $&$0.149 $&$27.9 $&$0.357  $&$0.011 $&$0.368$
&$0.031$&$0.534 $&$0.231$&$0.766 $&$    0.433$\\
P3&$0.005$&$0.071$&$0.076$&$2.70$&$0.179$&$0.012$&$0.191$&$0.068$
&$0.281 $&$0.121$&$0.402$&$0.431$\\
\hline\end{tabular}
\end{table}

We are aware that the OME models predict a too large and negative
asymmetry parameter $a_\Lambda$  in
$^5_\Lambda$He~\cite{Sa02,Pa02,Al05,Ba05,Ba07,Kr08}, and also that
there are two recent proposals to bring this value into agreement with
experiments by going beyond the OME model and incorporating   new
scalar-isoscalar terms. Namely, Chumillas \etal~\cite{Ch08} have
pointed out  that these new terms come from the exchange of
correlated $2\pi$ coupled to $\sigma$, plus uncorrelated $2\pi$ exchanges,
while Itonaga \etal~\cite{It08} had  to invoke  the axial-vector $a_1$
meson to reproduce the data for $a_\Lambda$
in $^5_\Lambda$He.
The $2\pi$-exchange potentials are rather cumbersome, and it is somewhat 
controversial to which extent these new mechanisms alter the transition rates. 
The first group ~\cite{Ch08} affirms that they leave them  basically unaltered, 
as seen from the results T7 and T7' in Table \ref{Table1}.
Yet, the second group~\cite{It08} asserts that they, not only bring the 
asymmetry parameter $a_\Lambda$ into agreement with recent measurement, but
improve also the $\Gamma_n/\Gamma_p$ ratio such as to become well
comparable to the experimental data. Anyhow,  in no one of these
works are discussed the transition rates in $^{4}_\Lambda$He and
$^{4}_\Lambda$H.
The solution  for the  $a_\Lambda$ puzzle might appear also from the
experimental side,  as has  occurred in the case of the $n/p$ branching ratio.

The transition probability densities $S_{N}(E)$, $S_{nN}(E)$, and
$S_{nN}(\cos\theta)$
contain the same dynamics, \ie the same NME's, but involve different 
phase-space kinematics for each case.
In particular, the proton spectrum
$S_{p}(E)$ is related with the expected number
of protons  $d{\rm N}_p(E)$   detected within the energy
interval $dE$ through the relation
\be
\frac{d{\rm N}_p(E)}{dE}=C_p(E)S_p(E),
\label{25}\ee
 where
$C_p(E)$ depends on the proton experimental environment and
includes all quantities and effects not considered in $S_p(E)$, 
such as  the  number of produced hypernuclei, 
the detection efficiency and acceptance, \etc.
In experiment E788, after correction for
acceptance, the remaining $C_p(E)$ factor is approximately energy-independent 
in the region beyond the detection threshold, $E^0_p$~\cite{Pa08}.
In what follows, we will always compare our predictions
with the experimental spectra that have been corrected for
acceptance and take into account the detection threshold. 
Thus we can write, for the expected number of detected protons above 
this threshold,
  \be
\bar{{\rm N}}_{p}=\int_{E_p^0}^{E_p^{max}} 
\frac{d{\rm N}_{p}(E)}{dE}dE=
\bar{C}_{p}\int_{E_p^0}^{E_p^{max}} S_{p}(E)dE=
\bar{C}_{p}\bar{\Gamma}_p.
\label{26a}\ee
 This allows us to rewrite \rf{25} in the form%
\footnote{A similar expression is valid for  the $\beta$-decay
 strength function (see, for instance, \cite[Eq. (5)]{Wi06}).}
 \be
\frac{d{\rm N}_p(E)}{dE}={\bar{{\rm N}}_p}\frac{S_p(E)}{\bar{\Gamma}_p} 
\qquad (E > E_p^0).
\label{27a}\ee
The  spectrum  $S_p(E)$ is normalized to the experimental one
by replacing $\bar{{\rm N}}_p$ in \rf{27a} with the 
acceptance-corrected number of actually observed protons,
\be
\bar{{\rm N}}_p^{exp}=\sum_{i=1}^{m} \Delta {\rm N}_p^{exp}(E_i),
\label{28a}\ee
where $\Delta {\rm N}_p^{exp}(E_i)$ is the acceptance-corrected number of 
protons  measured at energy $ E_i$ within a fixed energy bin $\Delta E_p$, 
and $m$ is the number of bins beyond the detection threshold.
Thus, the quantity that we have to confront with data is
%
 \be
\Delta {\rm N}_p(E)= 
{\bar{\rm N}}_p^{exp}\Delta E_p\frac{S_p(E)}{\bar{\Gamma}_p},
 \label{30}\ee
where the barred  symbols ($\bar{\rm N}_p^{exp}=4546$, and
$\bar{\Gamma}_p=  0.168$) indicate that the proton threshold  $E^0_p= 40$  MeV 
\cite{Pa08} has been considered in the numerical evaluation of the 
corresponding quantities.
In contrast to  $\Delta {\rm N}_p^{exp}(E_i)$, $\Delta {\rm N}_p(E)$
is a continuous function of $E$.

\begin{figure}[h]
\includegraphics[width=1.0\linewidth]{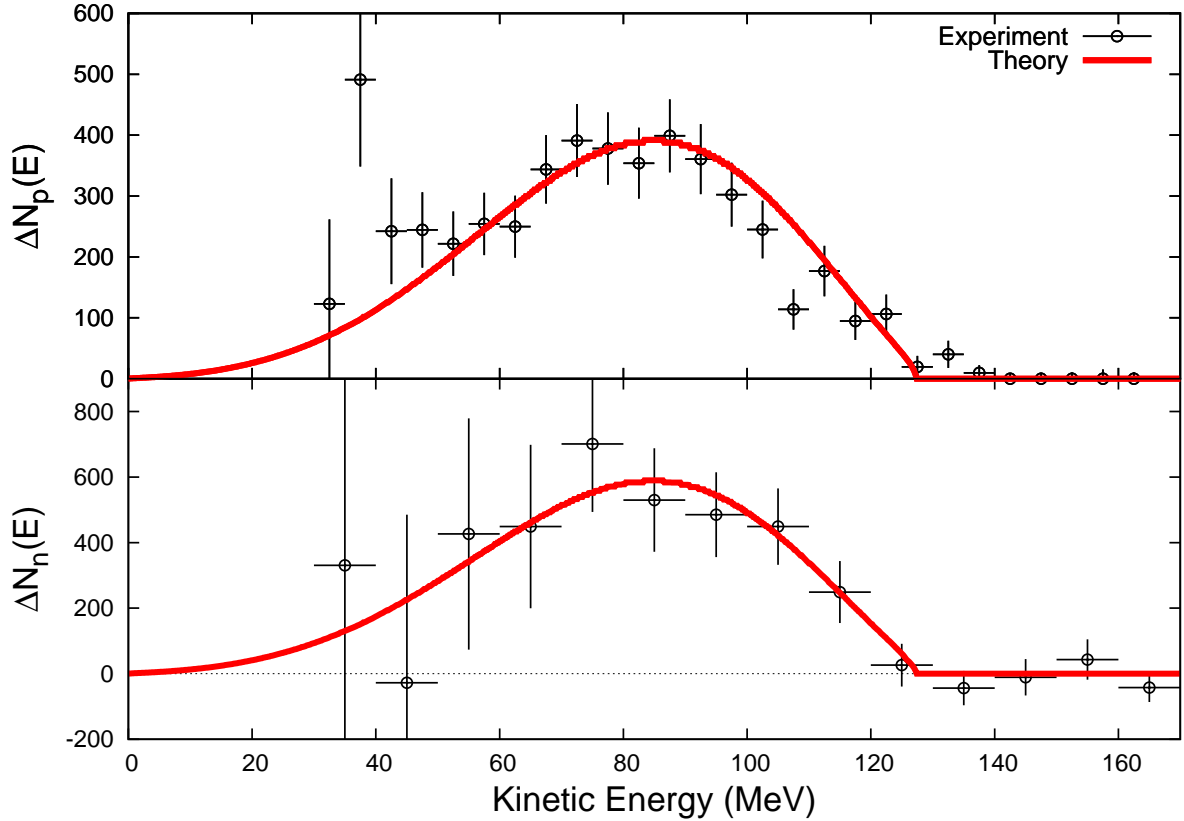}
\caption{\label{Figure2} Comparison between the experimental and theoretical 
kinetic energy spectra for protons (upper panel) and neutrons (lower panel). 
The data are acceptance corrected~\cite{Pa08},
and the calculated results are obtained from Eqs. \rf{30} and \rf{31}.}
\end{figure}

As the one-proton (one-neutron) induced decay prompts the emission of an $np$
($nn$) pair, one has in the same way for the one-neutron spectrum
\be
\Delta {\rm N}_n(E)= {\bar{\rm N}}_n^{exp}\Delta E_n 
\frac{S_p(E)+2S_n(E)}{\bar{\Gamma}_p+2\bar{\Gamma}_n}.
 \label{31}\ee
Here, ${\bar{\rm N}}_n^{exp}=3565$, and  $\bar{\Gamma}_p+2\bar{\Gamma}_n=0.198$
have been  evaluated with a neutron threshold of $ 30$  MeV \cite{Pa08}.
In Figure \ref{Figure2}, our results are compared  with the  measurements of
Parker \etal~\cite{Pa07}.

A similar, but somewhat different, procedure is followed for the coincidence 
spectra. The main difference arises from the fact that 
the angular-correlation spectra, $\Delta {\rm N}_{nN}^{exp}(\cos\theta_i)$, 
as well as the  kinetic energy sum data, $\Delta {\rm N}_{nN}^{exp}(E_i)$,
besides being  acceptance-corrected, were measured  with  detection thresholds  
of $30$ MeV for both neutrons and protons. 
More, 
in the selection of the kinetic energy sum data it
was also applied an angular cut of 
$\cos\theta_{nN}<-0.5$.
In order to make the presentation simple,  
the observables that comprise only the energy cuts, and those that
include both the energy and the angular cuts,
will be indicated by putting, respectively,  a tilde and a hat over the 
corresponding symbols.

\begin{figure}
\includegraphics[width=1.0\linewidth]{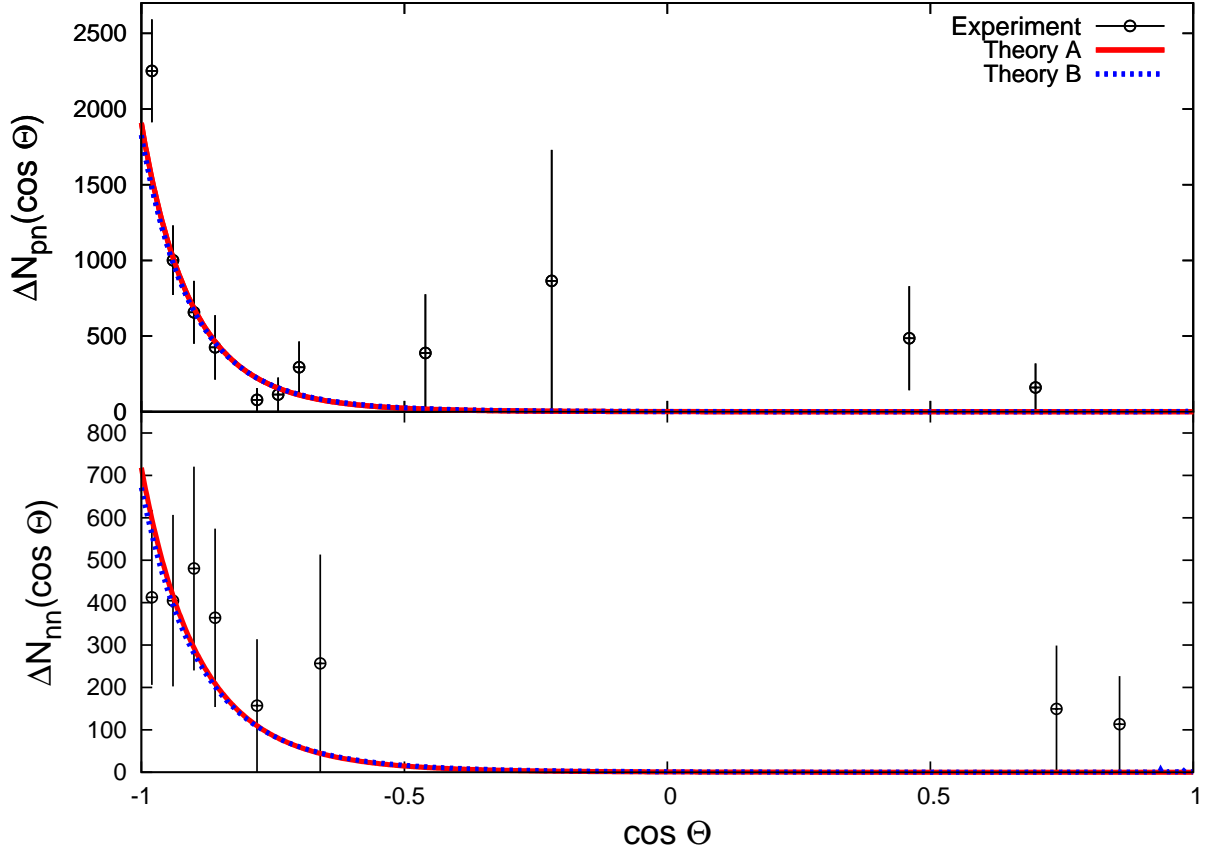}
\caption{\label{Figure3} Comparison between experimental opening angle 
correlations for proton-neutron (upper panel) and neutron-neutron (lower panel) 
pairs. The data $\widetilde{\Delta {\rm N}}_{nN}^{exp}(\cos\theta_i)$ are 
acceptance corrected and do not  contain
events with  $E_N<30$ MeV~\cite{Pa08}.
The theoretical results are obtained from  Eq. \rf{33},  
with $\widehat{{\rm N}}_{nN}^{exp}$
only containing  events with  $\cos\theta_{nN}<-0.5$.
Two cases are presented: 1) Theory A, where
both the angular and the single kinetic energy cuts are taken into account, 
and 2) Theory B,
where the cuts are not considered in the calculations.}
\end{figure}

\begin{figure}[h]
\includegraphics[width=1.0\linewidth]{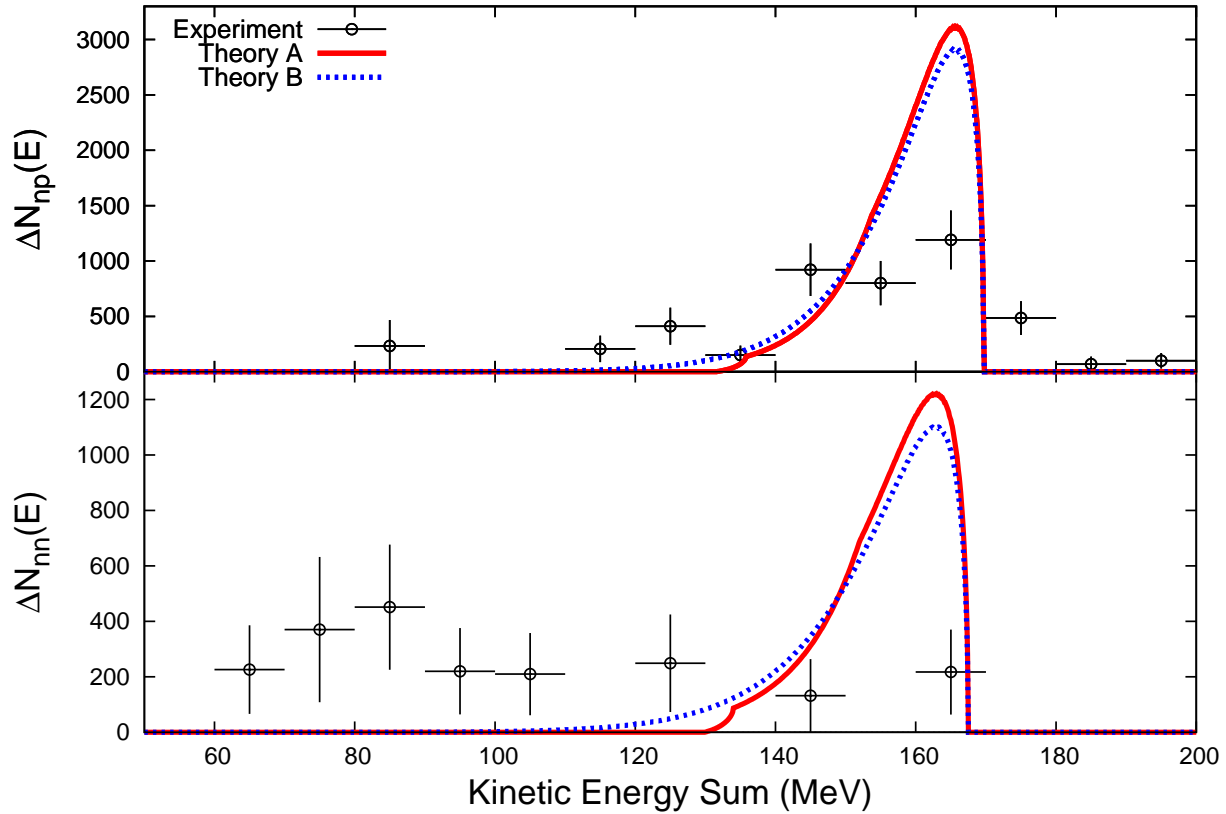}
\caption{\label{Figure4} Comparison between experimental kinetic energy sum 
spectra for proton-neutron (upper panel) and neutron-neutron (lower panel) 
pairs.
The data $\widehat{\Delta {\rm N}}_{nN}^{exp}(E_i)$ are acceptance corrected 
and only contain events with  $E_N>30$ MeV and  
$\cos\theta_{nN}<-0.5$~\cite{Pa08}.
The theoretical results are obtained from  Eq. \rf{34},  and
two cases are shown: 1) Theory A, where
both cuts are taken into account, and 2) Theory B,
where the cuts are not considered in the calculations.}
\end{figure}

Thus, the number of $nN$ pairs measured in coincidence can be expressed as
\be
\widehat{{\rm N}}_{nN}^{exp}
=\sum_{i=1}^{k} \widetilde{\Delta {\rm N}}_{nN}^{exp}(\cos\theta_i)=
\sum_{i=1}^{l} {\widehat{\Delta{\rm N}}}_{nN}^{exp}(E_i),
\label{32}\ee
where  the angular bins with $\cos\theta_i>-0.5$ are excluded from the first 
summation.
The $\widetilde{\Delta {\rm N}}_{nN}^{exp}(\cos\theta_i)$ and 
${\widehat{\Delta{\rm N}}}_{nN}^{exp}(E_i)$ data
should be compared, respectively, with
\be
\widetilde{\Delta {\rm N}}_{nN}(\cos\theta)=
\widehat{{\rm N}}_{nN}^{exp}\Delta\cos\theta_{nN}
\frac{\widetilde{S}_{nN}(\cos\theta)}{\widehat{{\Gamma}}_N},
\label{33}\ee
and
\be
\widehat{\Delta {\rm N}}_{nN}(E)=\widehat{{\rm N}}_{nN}^{exp}\Delta E_{nN} \frac{\widehat{S}_{nN}(E)}{\widehat{{\Gamma}}_N}.
\label{34}\ee
Here, from Ref.~\cite{Pa08}  $\widehat{{\rm N}}_{np}^{exp}=4821$, 
$\widehat{{\rm N}}_{nn}^{exp}=2075$,
$\Delta\cos\theta_{nN}=0.04$ and $\Delta E_{nN}=10$ MeV, 
while
$\widehat{\Gamma}_p=0.1709$ and 
$\widehat{\Gamma}_n=0.0113$. 
These  results (Theory A) are compared with the E788 data in 
Figures \ref{Figure3} and \ref{Figure4}.
For completeness, in the same figures are also shown the results for
$\widetilde{S}_{nN}(\cos\theta)\go{S}_{nN}(\cos\theta)$, 
$\widehat{S}_{nN}(E)\go {S}_{nN}(E)$ and 
$\widehat{\Gamma}_N\go {\Gamma}_N$,
\ie when no energy and angular
cuts are considered in the theoretical evaluation, and ${\Gamma}_p=0.1793$ and
${\Gamma}_n=0.0122$  (Theory B).

We conclude that the overall agreement between the  measurements of
Parker \etal~\cite{Pa07} and the present calculations is quite satisfactory, 
although we are not considering contributions coming from the two-body 
induced decay, $\Lambda NN\go nNN $,  nor from the rescattering of the nucleons 
produced in the one-body induced decay, $\Lambda N\go nN$.
However, before ending the discussion we would like to point out that:
\bnu
\item As expected,  the theoretical spectrum $\Delta {\rm N}_{p}(E)$,
shown in the upper panel of  Figure \ref{Figure2}, is
peaked around $85$ MeV, corresponding to the half of the $Q$-value 
$\Delta_p=170$ MeV.
Yet, as the  single kinetic energy reaches rather abruptly its maximum value 
$E^{max}_p=127$ MeV  (see Eq. \rf{13}), the proton spectrum shape is  not 
exactly  that of a symmetric inverted bell.
Something quite analogous  happens in the case of neutrons, as can be seen in 
the lower panel of  Figure \ref{Figure2}. 
The experimental data seem to behave in the same way.
To some extent, this   behavior of $\Delta {\rm N}_{p}(E)$ and  
$\Delta {\rm N}_{n}(E)$  is akin to the   behavior  of the 
$\Delta {\rm N}_{nN}(E)$, which  suddenly  collapse  at the Q-values. 

\item There are no data at really low energies for the
proton case which would allow to exclude the FSI effects for sure, and
the neutron data for low energies are afflicted by large error
bars. However, there is no need to invoke these effects, nor those of 
two-nucleon induced NMWD, to explain the data, as occurs in the proton 
spectrum of $^{5}_\Lambda$He~\cite{Ag08}. This hints at a new puzzle in 
the NMWD, but it is difficult to discern whether it is of experimental or 
theoretical nature.

 \item The  calculated spectra  $\widetilde{\Delta {\rm N}}_{np}(\cos\theta)$ 
shown in the upper panel of Figure \ref{Figure3},
are  strongly peaked near $\theta=180^o$,
which agrees with data fairly well. However, while it is found experimentally 
that $28 \%$ of events occur  at  opening angles less than $120^o$, 
theoretically we get that only $\lsim 2 \%$ of events appear in this angular 
region. We find no explanation for this discrepancy. 
Nevertheless, the  fact that not all events are concentrated at $\theta=180^o$, 
is not necessarily indicative of the  contributions coming from the  FSI or the 
$\Lambda NN\go nNN $ decay, as suggested  in Ref. \cite{Pa07}.

\item
The calculated angular correlation 
$\widetilde{\Delta {\rm N}}_{nn}(\cos\theta)$,  shown in
the lower panel of Figure \ref{Figure3}, is quite similar to that of the
$pn$ pair; that is, its   back-to-back peak is very pronounced. 
This  behavior  is not exhibited by the experimental distribution. 
In addition, while  $11 \%$ of events are found experimentally
for $\cos\theta \ge -0.5$, in the calculation only $\lsim 3 \%$ of them appear 
at these angles.
We feel however that, because of the poor statistics and large experimental 
errors, one should not attribute major importance to such  disagreements.

\item 
Both calculated kinetic energy sum distributions  
$\widehat{\Delta {\rm N}}_{nN}(E)$,  shown in
Figure \ref{Figure4}, present a bump at $\approx 160$ MeV, with a
width of $\approx 30$ MeV, which for protons agrees fairly well with
the experiment.
We would like to stress once more that the
spreading in strength here is totally normal even for a purely 
one-nucleon induced decay. 
The kink at  $\approx 130$ MeV within the Theory A comes from the angular cut, 
and from this one can realize  that the ${nN}$ kinetic energy sum spectra  
below this energy are correlated with the angular coincidence spectra 
$\widetilde{\Delta {\rm N}}_{nN}(\cos\theta<-0.5)$.
The bump  observed  in the experimental $\widehat{\Delta {\rm N}}_{nn}(E)$ 
spectrum at $\approx 90$ MeV is not reproduced by the theory, 
which may be indicative of  
$nn$ coincidences originated from sources other than  $\Lambda n $ decays, 
as already suggested in Ref.~\cite{Pa07}. 
Another source for the difference bewteen our model
calculation and the data may be traced to $np$ and $nn$ final state
interactions. Whereas in the former the intensity of this interaction
is reduced owing to the Coulomb repulsion felt by the proton, in
the latter the two neutrons may interact strongly and thus shift the
peak to lower kinetic energy sum. 
 \enu

In summary,  to comprehend the recent   measurements in $^4_\Lambda$He,
we have outlined  for the one-nucleon induced NMWD spectra
a simple theoretical framework based on  the IPSM.
Once normalized to the transition rate,  all the spectra are tailored 
basically by the kinematics of
the corresponding phase space, depending very
weakly on the dynamics 
governing the $\Lambda N \to nN$ transition proper.
As a matter of fact, although not shown here, the normalized spectra calculated 
with PSVE model are, for all pratical purposes, identical  
to those using the SPKE model, which we have amply discussed.
In spite of the simplicity of the approach, 
a good agreement with data is obtained.
This might indicate that, neither  the FSI, nor the two-nucleon induced
decay processes play a significant role 
in the $s$-shell, at least not for $^4_\Lambda$He.
As a byproduct  we have found that 
the $\pi + K$ exchange potential
with soft cutoffs (SPKE) is capable of accounting for the experimental values 
related to $\Gamma_p$ and $\Gamma_n$
in all three $^4_\Lambda$H, $^4_\Lambda$He, and $^5_\Lambda$He hypernuclei.
This potential  is not very different from the PKE model used by 
Sasaki \etal~\cite{Sa02}.

\begin{acknowledgments}
This work was partly supported by the Brazilian agencies FAPESP and CNPq,
 and by the Argentinian agency
CONICET under contract PIP 6159. MSH is the
2007/2008 Martin Gutzwiller Fellow at the Max-Planck-Institute for the
Physics of Complex Systems-Dresden.
We would like to thank G. Garbarino for very helpful discussions.
\end{acknowledgments}

\end{document}